\documentclass[ aip, rsi, amsmath,amssymb,  reprint,]{revtex4-1}%
\usepackage{graphicx}
\usepackage{dcolumn}
\usepackage{bm}%
\usepackage{amsmath}%
\usepackage{amsfonts}%
\usepackage{amssymb}
\providecommand{\U}[1]{\protect\rule{.1in}{.1in}}
\begin{document}

\preprint{AIP/123-QED}

\title[]{Surface effects on nitrogen vacancy centers neutralization in diamond}

\author{Arthur N. Newell}
\affiliation{Department of Physics and Engineering Physics, Delaware State University, Dover, Delaware 19901, USA}
\affiliation{ITAMP, Harvard-Smithsonian Center for Astrophysics, Cambridge,
Massachusetts 02138, USA}%
\author{Dontray A. Dowdell}%
\affiliation{Department of Physics and Engineering Physics, Delaware State University, Dover, Delaware 19901, USA}
\affiliation{ITAMP, Harvard-Smithsonian Center for Astrophysics, Cambridge,
Massachusetts 02138, USA}%

\author{D. H. Santamore}
 \email{dsantamore@desu.edu.}
\affiliation{Department of Physics and Engineering Physics, Delaware State University, Dover, Delaware 19901, USA}
\affiliation{ITAMP, Harvard-Smithsonian Center for Astrophysics, Cambridge,
Massachusetts 02138, USA}%

\date{\today}

\begin{abstract}
The performance of nitrogen vacancy (NV$^{-}$) based magnetic sensors strongly
depends on the stability of nitrogen vacancy centers near the diamond surface.
The sensitivity of magnetic field detection is diminished as the NV$^{-}$
turns into the neutralized charge state NV$^{0}$. We investigate the
neutralization of NV$^{-}$ and calculate the ratio of NV$^{0}$ to total NV
(NV$^{-}$+NV$^{0}$) caused by a hydrogen terminated diamond with a surface water
layer. We find that NV$^{-}$ neutralization exhibits two distinct regions:
near the surface, where the NV$^{-}$ is completely neutralized, and in the
bulk, where the\ neutralization ratio is inversely proportional to depth
following the electrostatic force law. In addition, small changes in
concentration can lead to large differences in neutralization behavior. This
phenomenon allows one to carefully control the concentration to decrease the
NV$^{-}$ neutralization. The presence of nitrogen dopant greatly reduces
NV$^{-}$ neutralization as the nitrogen ionizes in preference to 
NV$^{-}$ neutralization at the same depth. The water layer pH also affects
neutralization. If the pH is very low due to cleaning agent residue, then we
see a change in the band bending and the reduction of the $2$-dimensional hole
gas (DHG) region. Finally, we find that dissolved carbon dioxide resulting
from direct contact with the atmosphere at room temperature hardly affects the
NV$^{-}$ neutralization.

\end{abstract}
\maketitle

\preprint{AIP/123-QED}

\affiliation{Department of Physics and Engineering Physics, Delaware State University, Dover, Delaware 19901, USA}
\affiliation{ITAMP, Harvard-Smithsonian Center for Astrophysics, Cambridge,
Massachusetts 02138, USA}

\affiliation{Department of Physics and Engineering Physics, Delaware State University, Dover, Delaware 19901, USA}
\affiliation{ITAMP, Harvard-Smithsonian Center for Astrophysics, Cambridge,
Massachusetts 02138, USA}

\affiliation{Department of Physics and Engineering Physics, Delaware State University, Dover, Delaware 19901, USA}
\affiliation{ITAMP, Harvard-Smithsonian Center for Astrophysics, Cambridge,
Massachusetts 02138, USA}



\section{Introduction\label{Introduction}}

Nitrogen vacancy (NV$^{-}$) center diamonds have recently become the subject
of intense research for emergent quantum technologies such as metrology,
including nanoscale sensing
\cite{WrachtrupNat09,Degen08,LukinJakeNat08,WrachtrupNat08,LukinRon08,AwschalomAPL10,LukinChamPhys10,WrachtrupRevSciInst10,MerilesAPL10,HollenbPRB10,LiuNatNano11}%
; magnetometry
\cite{HellonNatNano11,CSPMagRes02,LukinRon14,DeganAPL14,HollenPRB14}; and
single-spin NMR \cite{HollenPRB14,JelezkoNatCom14}. Since NV$^{-}$ spin is
extremely sensitive to changes in the magnetic field, NV$^{-}$ center diamonds
can be used to detect paramagnetic objects optically and nondestructively.
Ideally, it is even possible to detect a single spin. NV$^{-}$-based devices
can be operated at room temperature and in air, which is a great advantage for
biological applications, because specimens will not be destroyed by the
scanning process. In recent years, considerable efforts have been made to use
NV$^{-}$ center diamonds for ultra-sensitive nanoscale magnetic sensor
heads\cite{Degen08,WrachtrupNat08,MitaPRB96} and sensor
arrays\cite{WrachtrupRevSciInst10,Fuchs10,HarvardNat13}. These sensors take
advantage of the long-lived spin state of single NV centers to detect small
magnetic fields \cite{WrachtrupNat08,LukinRon08}. Many nanoscale imaging
experiments using an NV center have provided magnetic images of various
objects, including disk drive media\cite{Harvard12,JacquesAPL12}, magnetic
vortices\cite{JacquesNatcom13}, a single electron spin\cite{HarvardNatphys13}
and magnetotactic bacteria\cite{HarvardNat13}. It should be emphasized that
magnetic sensors rely on an NV center having an extra electron (NV$^{-}$). The
neutralized form, NV$^{0}$, is not useful as a magnetic field sensor.

To detect a magnetic field in a sample, either the sample is moved over the
NV$^{-}$ center embedded in the bulk diamond, or an NV$^{-}$ embedded
cantilever tip scans over the sample, depending on whether the NV$^{-}$ center
is implanted at a shallow depth in the substrate or the NV$^{-}$ center is
located close to the apex of a tip. In both cases, it is very important to
maintain a sufficiently strong dipole-dipole coupling between the NV$^{-}$
center and the target spin for detection. The distance between the NV$^{-}$
center and the sample surface must be $10$ \textrm{nm} or less, and ideally
less than $5$ $\mathrm{nm}$. However, shallow NV$^{-}$ centers ($<5$
$\mathrm{nm}$) tend to become neutralized to NV$^{0}$, rendering magnetic
field detection impossible\cite{Degen13,Ania16}. The hydrogen surface
termination commonly used for semiconductor processing bends the band diagram
and creates a two dimensional hole gas ($2$-DHG) near the surface, which
accelerates the neutralization of NV$^{-}$
centers\cite{Wrach12,Wrach11,Fu10,Kucka11}. Other terminations, such as oxygen
and fluorine, do not have the energy band bending problem. However, oxygen
termination has its own problem. How oxygen bonds with carbon affects the band
diagram drastically. If a C-OH bond is formed, the surface interface has the
same problem with the hydrogen termination case causing the band bending and
accelerating the neutralization of NV$^{-}$ centers. On the other hand,
fluorine termination is reported to encounter severe blinking or
bleaching\cite{Degen13,Bradac13}. The detailed analyses about different
surface terminations and the bond structures are discussed by Kaviani et al.\cite{Gali14}. In
addition, NV centers interacting with paramagnetic contaminants such as water
molecules on the surface can decrease the sensitivity of a
magnetometer\cite{Penn02,Harvard15,DegenPRL14,HollenPRB14}. While experimental
progress in achieving ultra-high precision devices has been quite impressive,
the theory of bulk impurity interactions with NV centers and surface effects
is not well understood.

In this paper, we investigate the surface effects of NV$^{-}$ neutralization
in diamond. Sec.\ \ref{Sec_method} introduces the physical model (an NV$^{-}$
embedded diamond with a hydrogen terminated surface and a water layer) and
equations to be solved (the Poisson equation, the Schr\"{o}dinger-Poisson
equation, and the Poisson-Boltzmann equation), and describes the numerical
scheme to evaluate our model. Sec.\ \ref{Sec_results} shows our results and
discusses the physical implications. We examine the effects on NV$^{-}%
$\ neutralization with depth, concentration, addition of nitrogen, and pH of
the water layer. Finally, Sec.\ \ref{Sec_conclusion} concludes our discussions.

\section{Methods\label{Sec_method}}

\subsection{Model Structure\label{subsec_model}}

NV$^{-}$s are created by doping nitrogen in diamond through either chemical
vapor deposition or ion implantation. A nitrogen substitutes for a carbon atom
in a diamond unit cell. Then a vacancy next to the nitrogen in the unit cell
is formed by thermal annealing and an NV$^{-}$ is
formed\cite{Lang91,IakJPC01,HowTo12}. The typical yield of NV$^{-}$ from
nitrogen dopant is $0.1\mathrm{\%}-1\mathrm{\%}$; however, there are some
recent report of $10\mathrm{\%}$ and even $50\mathrm{\%}$\cite{Jayich16} at
the time of writing.

\begin{figure}[pth]
\begin{center}
\includegraphics[width=\columnwidth] {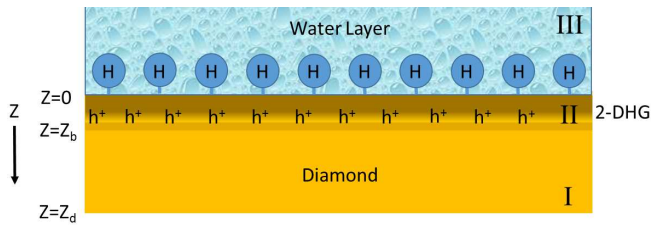}
\end{center}
\caption{Schematic representation of our model. The model consists of a bulk
diamond region (I), the hole $($h$^{+})$ accumulated region near the surface
(II), and the surface of the diamond with hydrogen termination and a water
layer (III).}%
\label{Fig1_schematic}%
\end{figure}

Figure \ref{Fig1_schematic} shows a schematic of our model. We study a
hydrogen-terminated $\left(  100\right)  $ diamond surface containing an
NV$^{-}$ center at room temperature and one atmospheric pressure as these are
the typical conditions at which most magnetic sensing NV$^{-}$ center diamond
devices operate. A thin water layer (usually a mono layer) condenses on the
surface from the water vapor in surrounding air. In addition, carbon dioxide
in the air may be dissolved in the water layer, changing the pH of the water.
The effect of pH changes on NV$^{-}$ neutralization will be described in
Sec.\ \ref{subsec_pH}.

The negative electron affinity (NEA) of hydrogen-terminated surface in vacuum
is reported to be $\chi=-1.3$ $\mathrm{eV}$\cite{LeyPRB01,LeyPSS00}. The NEA
leads to the depletion of electrons and the creation of a hole accumulation
layer (2-DHG), represented by (h$^{+}$) in Fig.\ \ref{Fig1_schematic}. The
depletion of electrons continues until the Fermi energy of the NV$^{-}$ center
diamond levels with the chemical potential of the atmosphere plus the water
layer. This causes the energy bands to bend. With the presence of a surface
water layer, the NEA can be even lower, further enhancing the bending.

There are three distinct regions that require different approaches to analyze
NV$^{-}$ neutralization (see Fig.\ \ref{Fig1_schematic}): Region I is the bulk
diamond away from the surface and the 2-DHG, Region II is the depletion region
including the 2-DHG, and Region III is the hydrogen terminated surface with
the water layer. In Fig.\ \ref{Fig1_schematic}, $z$ is the depth from the
surface. Two interfaces separating two regions are $z=0$ (the diamond and the
surface) $z=z_{b}$ (the depletion region and bulk diamond), and $z=z_{d}$ is
the edge of the diamond. We begin our analysis with Region II for mathematical
convenience as well as it being the most important region where the hole
accumulation and band bending occurs.

Region II is where the holes accumulate in the range $0<z<z_{b}$. The
depletion region has two forces acting on it: the electrostatic force and the
exchange and correlation force\cite{ApplePRB71}. The electrostatic force is
due to the charge density of other conduction electrons, the ionized donors,
and the dipole interaction with the surface water layer and contaminants
dissolved in water. The exchange and correlation forces result from other
conduction electrons, and between a conduction electron and the valence
electrons. The exchange and correlation forces are usually small; however,
they become important for analyzing band bending, the charge profile, the
density of states, and the subband structure. To account for these forces, we
use the Schr\"{o}dinger-Poisson equation. The Schr\"{o}dinger equation is used
for the states of the conduction holes, providing the charge density in terms
of the occupied states and the uniform-background charge density, while
Poisson's equation gives the potential in terms of the charge density
calculated by the Schr\"{o}dinger equation\cite{ApplePRB71}.

The Schr\"{o}dinger equation for the mobile states is
\begin{equation}
\left[  -\frac{\hbar^{2}}{2m_{n}}\frac{d^{2}}{dz^{2}}+V\left(  z\right)
\right]  \psi_{i}\left(  z\right)  =\varepsilon_{i}\psi_{i}\left(  z\right)  ,
\label{Eq_schrodinger}%
\end{equation}
where $z$ denotes the depth from the surface, $\psi_{i}\left(  z\right)  $ is
the hole sub-band $i$ state wave function with the eigenenergy $\varepsilon
_{i}$ and%
\begin{equation}
V\left(  z\right)  =q\Phi\left(  z\right)  +V_{h}.
\end{equation}
Here $q$ is the charge, $\Phi\left(  z\right)  $ is the electrostatic
potential, $V_{h}$ is the effective potential energy associated with the
interface discontinuity barrier (the difference between the pinned valence
band maximum at the interface and the Fermi energy) which is reported to be
$1.68$ \textrm{eV} \cite{Zrenner04}.

The boundary conditions are Dirichlet%
\begin{equation}
\psi_{i}\left(  0\right)  =\psi_{i}\left(  z_{b}\right)  =0,
\end{equation}
where $z=0$ is the interface. We set $z=z_{b}$ to be much deeper than the
$2$-DHG layer into the bulk diamond so that $z_{b}$ is well-away from the
$2$-DHG accumulation region, which allows us to contain the majority of the
band bending, as well as allows the wavefunction to be extended beyond the
depletion if needed. $m_{n}$ is the effective mass of holes, where $n$ runs
over the heavy hole (HH), light hole (LH) and split-off (SO) bands. Solutions
of Eq.\ (\ref{Eq_schrodinger}) are used to calculate the charge density in the
Poisson equation%

\begin{align}
\frac{d^{2}\Phi}{dz}  &  =-\frac{\rho\left(  z\right)  }{\epsilon}\nonumber\\
&  =-\frac{1}{\epsilon}2q\left[
{\displaystyle\sum\limits_{m,b}}
{\displaystyle\sum\limits_{i}}
f\left(  \varepsilon_{i}\right)  \psi_{i}^{\ast}\left(  z\right)  \psi
_{i}\left(  z\right)  \right]  +eN_{0}\nonumber\\
&  =\frac{2q}{\epsilon}%
{\displaystyle\sum\limits_{i}}
N_{i}\psi_{i}^{2}\left(  z\right)  +eN_{0}. \label{Eq_shroPoisson}%
\end{align}
Here $\Phi$ is the electrostatic potential, $\rho\left(  z\right)  $ is the
charge density, $\epsilon=\epsilon_{r}\epsilon_{0}=5.7\epsilon_{0}$ is the
permittivity of diamond ($\epsilon_{0}$ is the vacuum permittivity and
$\epsilon_{r}$ the relative permittivity.) The factor of two is the spin
degeneracy, $f\left(  \varepsilon_{i}\right)  $ is the probability of the
particle having energy $\varepsilon_{i}$ when the Fermi energy is at $E_{F}$,
and $m$ and $b$ stand for mobile and bound states. The second term $eN_{0}$ is
the uniform background charge density. The spatial density of particle $N_{i}$
is determined by the effective density of states and the Fermi-Dirac
distribution. We assume the dielectric constant is position independent.

The boundary conditions for the Poisson equation are%
\begin{align}
\Phi\left(  z_{b}\right)   &  =\Phi_{0},\\
-\left.  \frac{d\Phi}{dz}\right\vert _{z=0}  &  =E_{0},
\end{align}
where $E_{0}$ is the electric field at the surface of the diamond. The
Schr\"{o}dinger-Poisson equation can be solved only numerically.

Now considering region III ($z<0$), where electrolyte covers the surface, we
need to consider mobile charges such as molecules and ions rather than the
fixed ions of the solid. The Poisson equation is no longer applicable as
molecules and ions freely move about within the solution at finite
temperatures in thermal motion, even at equilibrium. The Poisson-Boltzmann
equation describes the distribution of the electric potential in the solution
in the direction normal to the charged surface.

Consider the situation where the interface ($z=0$) has the surface charge
density $\sigma$. If $\Phi\left(  \mathbf{z}\right)  $ is the electrostatic
potential, the charge density of the system is%
\begin{equation}
\rho\left(  z\right)  =\sigma\delta\left(  z\right)  +qn_{0}\exp\left[
-\frac{q\Phi}{k_{B}T}\right]  ,\label{Eq_electrolyte_rho}%
\end{equation}
where $k_{B}$ is the Boltzmann constant, $T$ is the temperature, and $n_{0}$
is the ion concentration very far from the interface). The freedom of movement
of ions in the electrolyte is accounted for by Boltzmann statistics.
Substituting Eq.\ \ref{Eq_electrolyte_rho} into the Poisson equation gives the
Poisson-Boltzmann equation,%
\begin{align}
\frac{d^{2}\Phi}{dz^{2}} &  =-\frac{1}{\epsilon}\left\{  \sigma\delta\left(
z\right)  +qn_{0}\exp\left[  -\frac{q\Phi}{k_{B}T}\right]  \right\}
,\label{Eq_PBeqn}\\
&  \Rightarrow\frac{d^{2}\Phi}{dz^{2}}+\frac{qn_{0}}{\epsilon}\exp\left[
-\frac{q\Phi}{k_{B}T}\right]  =-\frac{\sigma}{\epsilon}\delta\left(  z\right)
\end{align}
We have assumed that (a) the energy in the Boltzmann statistics depends only
on the electrostatic energy, (b) $\epsilon$ has no position dependence, (c)
point-like ions interact via their mean field, and (d) the water layer is
treated as a structureless continuum. With these assumptions, the
Poisson-Boltzmann equation becomes the nonlinear Guoy-Chapman model for the
electrolyte\cite{Gouy1910,Gouy21910,Chapman1913}. Although structural
properties of the liquid such as the sizes of ion and solvent and their
interactions can influence the ion energy, they are negligible since their
effects are small. In addition, we are interested in NV$^{-}$ neutralization
and not the dynamics of the electrolyte itself. If we want to know the details
of electrolyte distributions, then we would need to use molecular dynamics
simulations to introduce liquid structure rather than solving a non-linear
Gouy-Chapman model. However, molecular dynamics simulations are beyond the
scope of this paper since we are more concerned with the physics of the
diamond (Region I and II) than with the details of the electrolyte region
(Regions III).

One of the boundary conditions is obtained by integrating Eq.\ \ref{Eq_PBeqn}
over an infinitesimal interval around $z=0$. We also impose the boundary
condition that the electric field $E\left(  z\right)  \rightarrow0$ as
$z\rightarrow\infty$. If $q\Phi/k_{B}T\ll1$, then we can Taylor expand
$\exp\left[  -q\Phi\left(  \mathbf{r}\right)  /k_{B}T\right]  $ as
$\exp\left[  -q\Phi\left(  \mathbf{r}\right)  /k_{B}T\right]  \simeq
1-q\Phi\left(  \mathbf{r}\right)  /k_{B}T$, which is the Debye-H\"{u}ckel
approximation. The Debye-H\"{u}ckel approximation makes the Poisson-Boltzmann
equation analytically solvable. However, in our case, the potential calculated
from the numerical solution gives $q\Phi/k_{B}T>1$, so that the approximation
does not apply. We must therefore solve the full equation numerically.

Finally, Region I contains the majority of the bulk diamond and is far from
the surface and the $2$-DHG. This region can be solved by a simple Poisson
equation. The boundary conditions for the Poisson equation are
\begin{equation}
\Phi\left(  z_{b}\right)  =\Phi_{0},
\end{equation}%
\begin{equation}
-\left.  \frac{d\Phi}{dz}\right\vert _{z=z_{d}}=0.
\end{equation}

\subsection{The role of a surface water layer}

So far, we have always incorporated a water layer as an integral part of the
system in our model since water molecules are polar molecules.
Macroscopically, the water layer will function as an electric dipole layer on
the surface that will behave like the surface charge density $\sigma
$\cite{Jackson,Zangwill}. As a very rough estimate of the surface charge
density, let us model the monolayer of water as a dipole layer, two oppositely
charged surfaces with area charge densities $\pm\sigma$. The dipole moment of
a molecule is $\mathbf{p}=q_{\mathrm{eff}}\mathbf{b}$, where $b$ is the molecule's
length and $q_{\mathrm{eff}}$ is the effective charge. Assuming that each molecule
occupies an area $S$ of the interface, we can calculate a uniform electric
dipole moment per unit area $\tau$, where $\tau=p/S$. In the macroscopic
limit, $b\rightarrow0$ and $\sigma\rightarrow\infty$ with $\tau=\sigma b$ held
constant,
\begin{equation}
\frac{p}{S}=\frac{q_{\mathrm{eff}}b}{S}=\sigma b
\end{equation}
which leads to $\sigma=q_{\mathrm{eff}}/S$. The magnitude of the total dipole moment of
a water molecule is $p=6.1\times10^{-30}$ \textrm{C}$\cdot$\textrm{m}, and the
size is $0.275$ \textrm{nm}, the effective charge is $6.38\times10^{-20}%
$\textrm{C} and the uniform surface charge density is $8.4\times10^{-5}$
\textrm{C}$\cdot$\textrm{cm}$^{-2}$ ,which is comparable to the effective
surface charge that can cause the same band bending as the one caused by the
hydrogen termination. Therefore, the water layer contributes to the band
bending and affects the neutralization of NV$^{-}$, and we need to include the
surface water layer.

\subsection{Numerics\label{subsec_numerics}}

We calculate the NV$^{-}$ neutralization numerically since the equations we
have cannot be solved analytically. We are particularly interested in the
effects on NV$^{-}$ neutralization of a hydrogen terminated surface with a
water layer. We consider only nitrogen dopant and NV$^{-}$ and disregard other
bulk contaminants, such as $^{13}$C and B, in order to isolate the
neutralization caused by surface effects.

Diamond is an insulator and has an energy gap of $5.47$ \textrm{eV}. We set
the N$^{0/+}$ transition level to $1.7$ \textrm{eV} below the conduction band
minimum (CBM) and the NV$^{-/0}$ transition level to $2.67$ \textrm{eV} below
the conduction band minimum\cite{GaliPRB14}. Since many applications operate
NV$^{-}$ center diamond devices at room temperature, we take $T=298.15$
\textrm{K}.

The pH of the water layer is taken to be $7$ to simulate a layer of pure water
condensed on the surface. Realistically, the water layer can be slightly
acidic (pH of around $6$) when CO$_{2}$ is dissolved in the water when it is
in contact with air. We discuss that such changes in pH produce negligible
differences in the neutralization of NV$^{-}$ in Sec.\ \ref{subsec_pH}.

The surface of hydrogen terminated diamond is known to be hydrophobic. The
diamond and water layer are generated within a $1$-D model, as the
neutralization of the NV$^{-}$ depends primarily on the depth of the NV$^{-}$
from the surface.

As mentioned in Sec.\ \ref{subsec_model}, our model uses the nonlinear Poisson
equation in the bulk diamond (Region I), Schr\"{o}dinger-Poisson equation in
the depletion region (Region II), and the Poisson-Boltzman equation in the
water layer (Region III). We use the software package `nextnano$^{3}$' for
numerical evaluations\cite{Birner}. Nextnano$^{3}$ iteratively solves the
Poisson equation self-consistently within the discretized diamond and water
regions on a non-uniform grid by the finite-differences method. The
Schr\"{o}dinger-Poisson equation couples the Poisson equation to the
Schrodinger equation through the charge density in Eqs.\ (\ref{Eq_schrodinger}%
) and (\ref{Eq_shroPoisson}) within the diamond region. The Poisson-Boltzmann
equation is solved self-consistently using the same numerical scheme as in diamond.

\section{Results\label{Sec_results}}

\subsection{Band Diagram\label{subsec_band_diagram}}

\begin{figure}[pth]
\begin{center}
\includegraphics[width=\columnwidth] {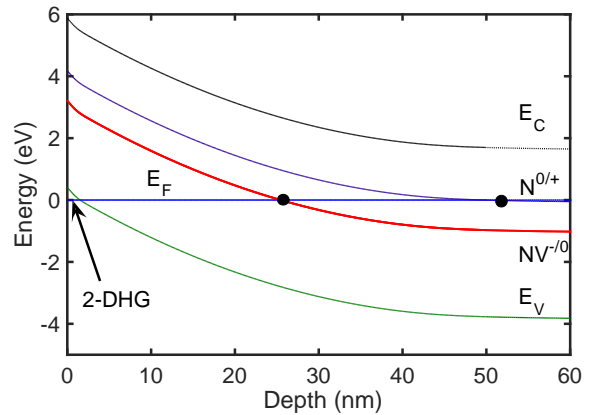}
\end{center}
\caption{Band diagram for diamond containing nitrogen and NV$^{-}$. All energy
levels calculated relative to a Fermi Level at $0$ \textrm{eV}. The region
where E$_{V}$ crosses E$_{F}$ is marked. This is the $2$-DHG region. The black
dots represent the depths up to which nitrogen and NV$^{-}$ are completely
neutralized.}%
\label{fig2_band}%
\end{figure}Figure \ref{fig2_band} shows the numerically evaluated band
diagram for a diamond containing a homogeneous doping profile of both nitrogen
and NV$^{-}$. The diamond is $150$ \textrm{nm }in depth and is doped with a
nitrogen concentration of $10^{18}$ \textrm{cm}$^{-3}$ and an NV$^{-}$
concentration of $10^{17}$ \textrm{cm}$^{-3}$. These particular concentrations
are used to simulate a $10\mathrm{\%}$ yield of NV$^{-}$ from nitrogen dopant.
The typical yield of NV$^{-}$ is $0.1\mathrm{\%}-10\mathrm{\%}$. The Fermi
level, $E_{F}$, is set to $0$ \textrm{eV} for convenience.

Hydrogen surface termination causes band-bending. A $2$-DHG as a result of
band-bending is shown in the figure. In Fig. \ref{fig2_band}, the $2$-DHG is
spread to the depth of around $1$ \textrm{nm }from the surface. The points
where the N$^{0/+}$ and NV$^{-/0}$ transition lines cross the Fermi level are
marked by black dots in Fig. \ref{fig2_band}. These points reflect
$100\mathrm{\%}$ ionization of N$^{0}$ to N$^{+}$\ and $100$$\mathrm{\%}$
neutralization of NV$^{-}$ to NV$^{0}$, respectively. The depletion region
depth is much smaller than the complete ionization/neutralization points
depths. However, the total charge of accumulated holes roughly equals that of
ionized N plus neutralized NV$^{-}$.

\subsection{Depth Dependence\label{subsec_depth}}

For an NV$^{-}$-based device to function as a magnetic sensor, the NV$^{-}$
must be placed near the surface. The magnetic field sensitivity is directly
related to the depth of an NV$^{-}$. However, shallow NV$^{-}$s tend to become
neutralized. Therefore, the key factor for device performance is the depth and
stability of NV$^{-}$; successful detection depends on maintaining the
dipole-dipole coupling strength by placing the NV$^{-}$ close to the surface
while, at the same time, keeping the NV$^{-}$ stable.

We first calculate the depth dependence of NV$^{-}$ neutralization for a
hydrogen terminated surface with only NV$^{-}$ and no nitrogen. This is done
to isolate the NV$^{-}$ neutralization process from nitrogen ionization
process since both presences result in donating electrons.

The doping profile used here is a Gaussian distribution where the magnitude of
the peak point and the standard deviation (straggle) are kept constant at each
depth. The standard deviation is set to a much smaller value than the depth
($0.5$ \textrm{nm}). We use this doping profile to accurately determine the
neutralization rate at each depth. In practice, such doping profile can be
produced by chemical vapor deposition or a low energy ion implantation at
shallow depths. The peak concentration is set at $10^{20}$ \textrm{cm}$^{-3}$.

\begin{figure}[pth]
\begin{center}
\includegraphics[width=\columnwidth] {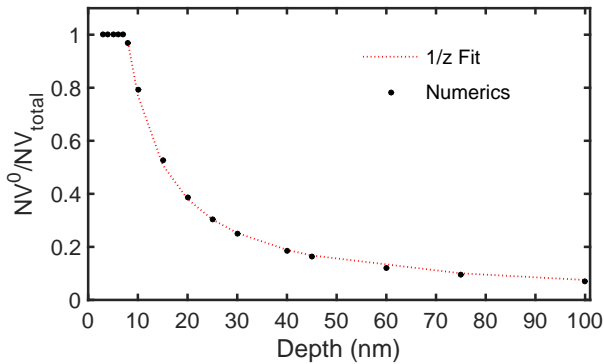}
\end{center}
\caption{Depth dependence of the NV$^{-}$ neutralization described by the
neutralization ratio of NV$^{-}$. The same Gaussian profile is used for the
NV$^{-}$ distribution to evaluate neutralization at each depth. The standard
deviation is $0.5$ \textrm{nm} and the peak concentration is set at $10^{20}$
\textrm{cm}$^{-3}$. Points represent values from the simulation; the dotted
line represents the $1/z$ fit. The neutralization ratio is taken as NV$^{0}%
/$NV$_{\mathrm{total}}$, where NV$^{0}$ is the concentration of NV$^{0}$ and
NV$_{total}$ is the total concentration of NV centers.}%
\label{fig3_depth}%
\end{figure}\qquad\qquad\qquad\qquad\

Figure \ref{fig3_depth} shows the neutralization ratio, NV$^{0}/$%
NV$_{\mathrm{total}}$, as a function of depth ranging from $3$ \textrm{nm} to
$100$ \textrm{nm}. There are two distinct regions: a plateau where all
NV$^{-}$s are completely neutralized to NV$^{0}$s and a monotonically
decreasing region, where the neutralization is inversely proportional to the
depth, $z$ (see Fig.\ \ref{fig3_depth}). The $1/z$ dependence indicates that
the major driving force in neutralization is electrostatic.

The transition from the plateau to the monotonically decreasing region occurs
when the potential difference generated by the accumulated holes becomes less
than the total energy required to neutralize all NV$^{-}$ to NV$^{0}$.

\subsection{Concentration Dependence\label{subsec_conc}}

\begin{figure}[pth]
\begin{center}
\includegraphics[width=\columnwidth] {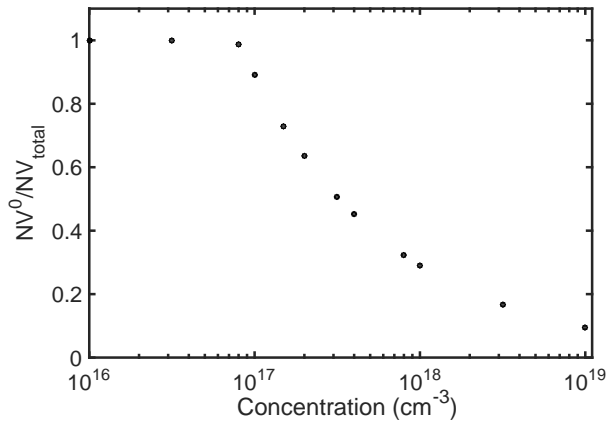}
\end{center}
\caption{Results of concentration dependence simulations. A homogeneous
NV\ profile is used for each concentration. Concentration is plotted on a
logarithmic scale. The neutralization ratio is shown as NV$^{0}/$%
NV$_{\mathrm{total}}$, where NV$_{total}$ is the total number of NV centers.}%
\label{fig4_Conc}%
\end{figure}

As seen in Sec.\ \ref{subsec_depth}, shallow depth promotes the neutralization
of NV$^{-}$. Since the amount of neutralization is governed by the
electrostatic force, one way to offset neutralization is to increase
concentration. Figure\ \ref{fig4_Conc} shows the neutralization ratio as a
function of NV$^{-}$ concentration. The NV$^{-}$ profile is taken to be
homogeneous throughout the diamond structure to remove any depth dependence.
Concentrations of NV$^{-}$ ranging from $10^{16}$ to $10^{19}$ \textrm{cm}%
$^{-3}$ are used to accommodate a recent experimental achievement of
$50\mathrm{\%}$ NV$^{-}$ conversion yield from doped nitrogen in a diamond
sample\cite{Jayich16}. The concentration of $10^{19}$ \textrm{cm}$^{-3}$ is
the maximum feasible doping concentration of nitrogen according to Element
Six$^{\mathrm{TM}}$.\cite{Element6}

\qquad

Figure \ref{fig4_Conc} shows the concentration dependence of NV$^{-}$
neutralization. As the concentration increases, the neutralization decreases.
Increasing concentration decreases the depth of the fully neutralized region.
Our calculation indicates that the total number of neutralized NV$^{0}$ is
less than the total number of accumulated hole created (the ratio of the
number of the neutralized NV$^{-}$s to the number of the accumulated holes is
$0.76$ to $1$). We account this to the relatively high neutralization energy
NV$^{0}$ ($2.67$ \textrm{eV}). In Sec.\ \ref{Subsec_Ns_effects}, when we add
nitrogen to the system, with ionization energy $1.7$ \textrm{eV}, the ratio of
the total number of the neutralized NV$^{-}$ plus the ionized nitrogen to the
number of the accumulated holes is $1$ to $1$.

\subsection{Effects of Concentration on Depth
Dependence\label{subsec_conc_depth}}

In Sec.\ \ref{subsec_depth}, \ the overall concentration defined by the peak
concentration and standard deviation of a Gaussian profile is kept the same at
each depth. We now investigate how the peak concentration of NV$^{-}$ affects
neutralization. We keep the same standard deviation of the Gaussian function
as in Sec.\ \ref{subsec_depth}, while varying the peak concentration of the
doping profile. We again assume that there is only NV$^{-}$, with no nitrogen.

\begin{figure}[pth]
\begin{center}
\includegraphics[width=\columnwidth] {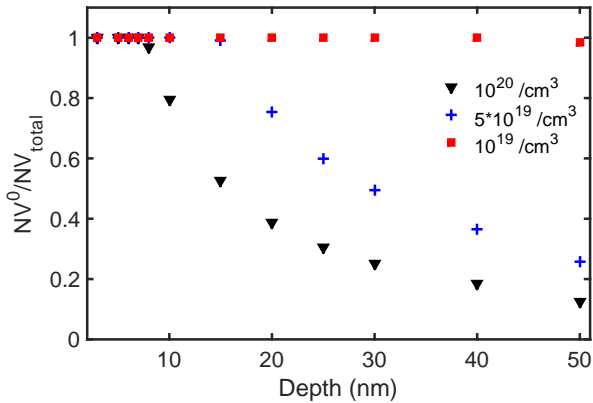}
\end{center}
\caption{Neutralization ratios of NV$^{0}/$NV$_{\mathrm{total}}$ for each
depth and peak concentration simulated. The diamond was modelled as being
doped with only NV$^{-}$. A Gaussian profile was used for the NV$^{-}$
distribution. The standard deviation was fixed at $0.5$ \textrm{nm} for each
profile, while varying the peak concentration.}%
\label{fig5_ConcAndDepth}%
\end{figure}

Figure\ \ref{fig5_ConcAndDepth} shows the neutralization ratio as a function
of depth at three slightly different concentrations. The figure indicates that
neutralization occurs from the nearest surface preferentially as expected. Our
results also suggest that slight changes in concentration produce drastic
effects. With a five-fold increase in concentration, from $10^{19}$
\textrm{cm}$^{-3}$ to $5\times10^{19}$ \textrm{cm}$^{-3}$, the neutralization
ratio changes from nearly complete neutralization to partial neutralization.
Increasing the concentration from $10^{19}$ \textrm{cm}$^{-3}$ to $10^{20}$
\textrm{cm}$^{-3}$ pushes the neutralization curve further down reducing the
neutralization ratio from $1$ to $0.79$ at a relatively shallow depth of $10$
\textrm{nm}. Therefore, one can control the stabilization of NV$^{-}$ at
shallow depth by carefully controlling the peak concentration of NV$^{-}$.

\subsection{Effects of Nitrogen\label{Subsec_Ns_effects}}

Having modelled both the depth and concentration dependence of NV$^{-}$
neutralization, we now add nitrogen to the system. Again we use a homogeneous
concentration profile for both nitrogen and NV$^{-}$. The nitrogen
concentration is fixed at $10^{18}$ \textrm{cm}$^{-3}$. We examine
neutralization with $0.1\mathrm{\%}$, $1\mathrm{\%}$ and $10\mathrm{\%}$
NV$^{-}$ yields from the nitrogen dopant.

\begin{figure}[pth]
\begin{center}
\includegraphics[width=\columnwidth] {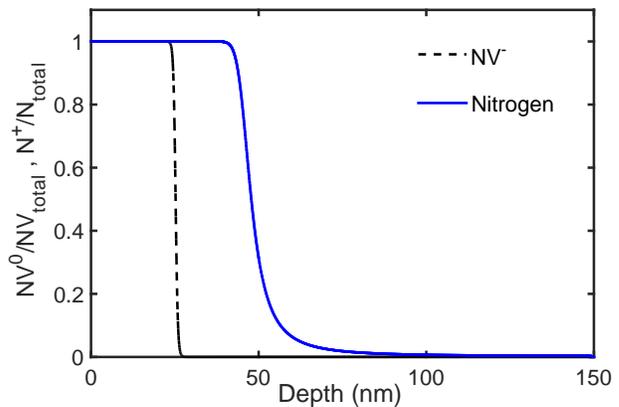}
\end{center}
\caption{Neutralization profile of nitrogen and NV$^{-}$. The neutralization
profile for NV centers is calculated as the ratio of NV$^{0}/$%
NV$_{\mathrm{total}}$ at each depth, where NV$^{0}$ is the total final
concentration of NV$^{0}$ and NV$_{\mathrm{total}}$ is the total initial
concentration of NV centers. The neutralization profile of nitrogen is
calculated as the ratio of N$^{+}/$N$_{total}$ at each depth, where N$^{+}$ is
the total final concentration of N$^{+}$ and N$_{\mathrm{total}}$ is the total
initial concentration of nitrogen.}%
\label{fig6_NandNV}%
\end{figure}

Figure \ref{fig6_NandNV} shows only a $1\%$ yield of NV$^{-}$ result,
\textit{i.e.}, at a density of $10^{16}$ \textrm{cm}$^{-3}$, for an easier
view. Even though both NV$^{-}$ and nitrogen are homogeneously distributed
throughout the diamond, one can see that nitrogen ionization is energetically
favored over NV$^{-}$ neutralization. As briefly mentioned in
Sec.\ \ref{subsec_numerics}, the ionization energy of nitrogen is $1.7$
\textrm{eV} below the conduction band maximum (CBM), while the neutralization
energy of NV$^{-}$ is $2.67$ \textrm{eV} below the CBM. Thus, the nitrogen
will become ionized more easily than the NV$^{-}$ will be neutralized. In
addition, because the ionization energy required for nitrogen is much larger
compared to typical semiconductor dopants ($\ll1$ \textrm{eV}), the nitrogen
dopant is hardly ionized at room temperature. As a result, there are enough
nitrogen atoms left to ionize in preference to NV$^{-}$ by the hydrogen
termination. Therefore, when nitrogen is introduced into a system with only
NV$^{-}$, the NV$^{-}$ neutralization is suppressed as long as the following
conditions are both met: (a) the potential difference created by the
electrostatic force, which depends on distance, cannot overcome the
neutralization energy and (b) there are still some nitrogen atoms left for ionization.

Finally, by having a high nitrogen concentration, one can decrease
neutralization at shallow depths. In principle, depths of less than $10$
\textrm{nm} for stable NV$^{-}$ are viable.

\subsection{Effects of pH\label{subsec_pH}}

\begin{figure}[pth]
\begin{center}
\includegraphics[width=\columnwidth] {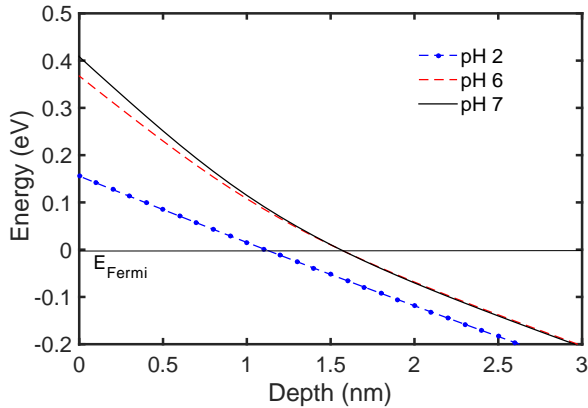}
\end{center}
\caption{Close-up of the band diagram around the $2$-DHG region for diamond
containing NV$^{-}$ at water pH $=7$ (solid lines), pH $=6$ (dashed lines) and
pH $=2$ (dotted lines) , where the difference between the band diagrams can be
seen.}%
\label{fig7_pH}%
\end{figure}

In each of our models, we have taken the water layer to have a pH of $7$. In
practice, the water layer is likely to take pH values near but not equal to
$7$: CO$_{2}$ from the air is dissolved in water can lower the pH of the water
layer to $6$. Here, we examine NV$^{-}$ neutralization at a different pH (see
Fig.\ \ref{fig7_pH}). The band diagram is shown with only the valence bands
and the Fermi energy level and around the $2$-DHG region (Fig.\ \ref{fig7_pH}),

The band structures are very similar for the systems with the water layer at
pH $6$ and pH $7$. So, the concentrations of holes accumulated near the
surface for both cases are also similar implying that neutralization of the
NV$^{-}$ would be similar. In fact, the neutralization ratio almost remains
the same in the range of pH $6-8$. Therefore, neutralization of NV$^{-}$ is
not affected by a pH change around pH $7$. However, much lower pH has a large
effect on band bending. In practice, much lower pH may be caused by a
incomplete removal of a surface cleaning agent, such as sulfuric acid from the
surface. For example, when the concentration $0.1$ \textrm{mol/L} of sulfuric
acid is left on the surface, the pH of the water layer could drop down to as
low as $2$. We see a change in the band bending and the drastic reduction of
the $2$-DHG region at pH $2$ (see Fig.\ \ref{fig7_pH}).

\subsection{Conclusion\label{Sec_conclusion}}

We have investigated the neutralization mechanisms of NV$^{-}$ centers in
diamond with a hydrogen terminated surface and a water layer on the surface.
The hydrogen terminated diamond forms a $2$-DHG right below the interface. The
depth of the $2$-DHG changes with the dopant concentrations and distributions.
The size of the $2$-DHG reflects the amount of accumulated holes. A thin water
layer condenses on the surface from the water vapor in the surrounding air.

The neutralization is quantified as the ratio of NV$^{0}$ at equilibrium to
the total number of NV originally implanted, NV$^{0}/$NV$_{\mathrm{total}}$.
The model is comprised of three regions, with appropriate equations used in
each case: the Poisson equation for the bulk diamond region, the
Schr\"{o}dinger-Poisson equation for the hole accumulated region, and the
Poisson-Boltzmann equation for the water layer.

We have calculated the neutralization ratio, NV$^{0}/$NV$_{\mathrm{total}}$ at
various depths. We find that the the NV$^{-}$ center neutralization is
primarily electrostatic. For high initial concentrations of NV centers, we
have found that NV$^{-}$ can remain stable even at depths of less than $10$
\textrm{nm}.

As the concentration increases, the neutralization decreases. The NV$^{0}%
$\ distribution as a function of depth shows that neutralization starts at
sites near the surface as expected. Increasing concentration decreases the
neutralization ratio. This is simply caused by the increase in the shear
number of available NV$^{-}$s as the concentration increases. However, when
the depth dependence is examined with various but close concentrations, our
results suggest that small changes in the initial concentration could lead to
large changes in the NV$^{-}$ neutralization. The results are useful to
control the stability of NV$^{-}$ at shallow depths.

The addition of nitrogen to the system shows that energetically, at a given
depth, nitrogen ionization is favored over NV$^{-}$ neutralization due to the
nitrogen ionization energy being lower than the NV$^{-}$\ neutralization energy.

Finally, we have shown that small variations in the pH of the water layer will
lead to insignificant changes in the $2$-DHG region within the diamond. Such
small changes mean that small variations in the pH of a water layer around pH
$7$ for a hydrogen terminated diamond will not affect the neutralization of
NV$^{-}$ centers within the diamond.

\begin{acknowledgments}
This work is supported by NSF DMR-1505641. We are grateful to ITAMP at
Harvard-Smithsonian Center for Astrophysics for hosting us during the summer,
where the majority of this work has been carried out. We thank Hossein
Sadeghpour, Mikhail Lukin and his group, and Ron Walsworth and his group for
useful discussions and valuable input.
\end{acknowledgments}

\nocite{*}

\end{document}